\documentclass[showpacs,preprintnumbers,amsmath,amssymb,twocolumn,superscriptaddress,prb]{revtex4}
\usepackage{times}
\usepackage{graphicx}
\usepackage{amsfonts}
\usepackage{amsmath, amsthm, amssymb}

\newcommand{\ve}[1]{\mathbf{#1}}

\newcommand{\vk}{\ve{k}} 
\newcommand{\vp}{\ve{p}} 
\newcommand{\vg}{\mathbf{v}_\text{g}} 

\newcommand{\e}[1]{\mathrm{e}^{#1}}

\newcommand{\pauli}{\boldsymbol{\hat{\sigma}}}
\def\i{\mathrm{i}} 

\newcommand{\ie}{\textit{i.e. }}
\newcommand{\eg}{\textit{e.g. }}
\newcommand{\etal}{\emph{et al.}}

\begin{document}
\title[Tunneling conductance in $s$- and $d$-wave superconductor-graphene
           junctions:  Extended Blonder-Tinkham-Klapwijk formalism
]{Tunneling conductance in $s$- and $d$-wave superconductor-graphene
           junctions:  Extended Blonder-Tinkham-Klapwijk formalism
}
\author{Jacob Linder}
\affiliation{Department of Physics, Norwegian University of
Science and Technology, N-7491 Trondheim, Norway}
\author{Asle Sudb{\o}}
\affiliation{Department of Physics, Norwegian University of
Science and Technology, N-7491 Trondheim, Norway}
\date{Received \today}
\begin{abstract}
We investigate the conductance spectra of a normal/superconductor graphene junction using the extended Blonder-Tinkham-Klapwijk formalism, considering pairing potentials that are both conventional (isotropic $s$-wave) and unconventional (anisotropic $d$-wave). In particular, we study the full crossover from normal to specular Andreev reflection without restricting ourselves to special limits and approximations, thus expanding results obtained in previous work. In addition, we investigate in detail how the conductance spectra are affected if it is possible to induce an unconventional pairing symmetry in graphene, for instance a $d$-wave order parameter. We also discuss the recently reported conductance-oscillations that take place in normal/superconductor graphene junctions, providing both analytical and numerical results.
\end{abstract}

\maketitle

\section{Introduction}
A key issue in understanding low-energy quantum transport at the interface of a non-superconducting and superconducting material, \eg a normal/superconductor (N/S) interface, is the process of Andreev reflection. Although the existence of a gap in the energy spectrum of a superconductor implies that no quasiparticle states may persist inside the superconductor for energies below that gap, physical transport of charge and spin is still possible at a N/S interface in this energy-regime if the incoming electron is reflected as a hole with opposite charge. The remaining charge is transferred to the superconductor in the form of a Cooper pair at Fermi level. The study of Andreev reflection and its signatures in experimentally observable quantities such as single-particle tunneling and the Josephson current has a long history (see \eg Ref.~\onlinecite{deutscher} and references therein). Only recently, however, has this field of research been the subject of investigation in \textit{graphene} N/S interfaces \cite{beenakker, beenakkerreview}. \\
\par
\begin{figure}[h!]
\centering
\resizebox{0.40\textwidth}{!}{
\includegraphics{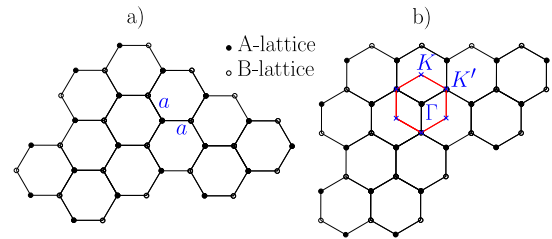}}
\caption{(Color online) a) Sketch of real-space lattice of graphene, consisting of two hexagonal sublattices A and B. The interatomic distance $a$ is equal between all lattice points. b) $\vk$-space (momentum-space) lattice of graphene, including the hexagonal Brillouin zone. Only two inequivalent points exist on the BZ boundary, termed $K$ and $K'$.}
\label{fig:honeycomb}
\end{figure}

Graphene is a monoatomic layer of graphite with a honeycomb lattice structure, as shown in Fig. \ref{fig:honeycomb}, and its recent experimental fabrication \cite{novoselov, zhang} has triggered a huge response in both the theoretical and experimental community over the last two years. The electronic properties of graphene display several intriguing features, such as the six-point Fermi surface and a Dirac-like energy dispersion, effectively leading to an energy-independent velocity and zero effective mass at the Fermi level. This obviously attracts the interest of the theorist, but graphene may also hold potential for technological applications due to its unique combination of a very robust carbon-based structural texture and its peculiar electronic features.\\
\par
\begin{figure}[h!]
\centering
\resizebox{0.30\textwidth}{!}{
\includegraphics{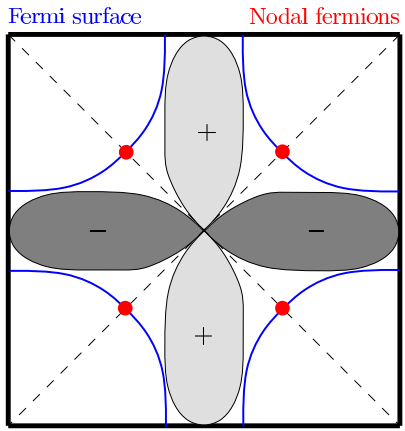}}
\caption{(Color online) Sketch of the Fermi surface and anisotropic $d$-wave gap as believed to be present 
in high-$T_c$ cuprate superconductors. The nodal fermions reside at the intersection of the nodal lines of 
the gap and the Fermi surface.}
\label{fig:nodal}
\end{figure}

Condensed matter systems with such `relativistic' electronic structure properties as graphene constitute 
fascinating examples of low-energy emergent symmetries; in this case, Lorentz-invariance. At half-filling, the Fermi level of graphene is exactly zero which renders the Fermi surface to be reduced to a six single points due to the linear intersection of the energy bands (see Fig. \ref{fig:dispersion} and \ref{fig:compare}). The linear dispersion relation is a decent approximation even for Fermi levels as high as 1 eV, such that the fermions in graphene behave like they are massless in the low-energy regime. The fact that the fermions around Fermi level obey a Dirac-like equation at half-filling introduces Lorentz-invariance as an emergent symmetry in the low-energy sector. Another example where Lorentz-invariance appears for low-energy excitations is in one-dimensional interacting fermion systems, where phenomena like breakdown of Fermi-liquid theory and spin-charge separation take place. When Lorentz-invariance emerges in the low-energy sector of higher-dimensional condensed matter systems, it 
is bound to attract much interest from a fundamental physics point of view. Another interesting feature of graphene 
are the nodal fermions that are present at the Fermi level at half-filling. When moving away from half-filling by 
doping, the excitations at the Fermi level are no longer nodal. The nodal fermions of graphene hold certain similarities to, 
but also important differences from, the nodal Dirac fermions appearing in the low-energy sector of the pseudogap 
phase of $d$-wave superconductors such as the high-$T_c$ cuprates. In contrast to graphene, the nodal fermions in 
the high-$T_c$ cuprates track the Fermi level when these systems are doped and thus represent a more robust feature than in graphene. For an illustration of the latter scenario, 
consider Fig. \ref{fig:nodal} which contains a sketch of the Fermi surface in the cuprates when including terms up 
to next-nearest neighbor hopping. The nodal lines of the $d_{x^2-y^2}$-gap intersect the Fermi surface at exactly 
four points, which permits the existence of nodal fermions at those points in $\vk$-space. However, in contrast to graphene, 
doping the system will in this case simply move the position of the Fermi arc with respect to the nodal line of 
the superconducting gap, such that the nodal fermions persist in the system \cite{QED3}. Nonetheless, the existence of
the Dirac cones in graphene represents an important example of emergent non-trivial symmetries at
long distances and low energies in higher (more than one) dimensional systems. 
\par
Although superconductivity does not appear intrinsically in graphene, it may be induced by means of the 
proximity effect by placing a superconducting metal electrode near a graphene layer 
\cite{heersche, kasumov, morpurgo, buitelaar, jariloo}. Recent theoretical work 
\cite{beenakker, sengupta} have considered coherent quantum transport in N/S and normal/insulator/superconductor (N/I/S) graphene junctions in 
the case where the pairing potential is isotropic, \ie $s$-wave superconductivity. However, the hexagonal 
symmetry of the graphene lattice permits, in principle, for unconventional order parameters such as $p$-wave 
or $d$-wave (characterized by a non-zero angular momentum of the Cooper pair). A complete classification of 
the possible pairing symmetries on a hexagonal lattice up to $f$-wave pairing ($l=3$) was 
given by Mazin and Johannes \cite{mazinjohannes}, with the result given in Tab. \ref{tab:symmetry}. We underline that the notation "insulator" in this context refers to a normal segment of graphene in which one experimentally induces an effective potential barrier $V_0$. As we shall see, such a potential barrier has dramatically different impact upon the transport properties in graphene as compared to the metallic counterpart.
\par
\begin{table}[h!]	
\centering{
\caption{List of all superconducting pairing states allowed for a hexagonal lattice up to $d$-wave pairing, 
adapted from Ref.~\onlinecite{mazinjohannes}. An orbital angular momentum quantum number $l=0,1,2$ is denoted 
$s, p, d$-wave, respectively. For the triplet states ($p$-wave), the order parameter has multiple components, 
and is conveniently represented as a vector $\mathbf{d}_\vk$.}
\label{tab:symmetry}
	\vspace{0.15in}
	\begin{tabular}{ccccc}
	  	 \hline
	  	 \hline
		 \textbf{Pairing}	\hspace{0.05in}	& \textbf{Type} \hspace{0.05in}&  \textbf{Pairing}	\hspace{0.1in}	& \textbf{Type} \\
	  	 \hline
	  	 1  & $s$ & $(k_y,-k_x,0)$ & $p$\\
	  	 $k_x^2+k_y^2$ & $s$ & $(0,0,k_z)$ & $p$ \\
	  	 $k_z^2$ & $s$ & $(k_x,k_y,0)$ & $p$ \\
	  	 $(0,0,k_x)$ & $p$ & $(k_x\pm\i k_y, \pm\i k_x - k_y,0)$ & $p$ \\
	  	 $(0,0,k_y)$ & $p$ & $(k_x\pm\i k_y)^2$ & $d$ \\
	  	 $(0,0,k_x\pm\i k_y)$ & $p$ & $k_xk_z$ & $d$ \\
	  	 $(k_z,0,0)$ & $p$ & $k_yk_z$ & $d$ \\
	  	 $(0,k_z,0)$ & $p$ & $(k_x\pm\i k_y)k_z$ & $d$ \\
	  	 $(k_z, \pm\i k_z,0)$ & $p$ & $k_x^2 - k_y^2$ & $d$ \\
	  	 $(k_y, k_x,0)$ & $p$ & $k_xk_y$ & $d$ \\
	  	 $(k_x,-k_y,0)$ & $p$ \\
	  	 \hline
	  	 \hline
	\end{tabular}}
\end{table}
\par
The intrinsic spin-orbit coupling in graphene is very weak, as dictated by the low value of the carbon atomic number, such that we will neglect it in this work. We will also disregard the electrostatic repulsion as mediated by the vector potential $\mathbf{A}$. At first sight, this might seem as an unphysical oversimplification since there is no metallic screening of the Coulomb interaction in graphene. In an ordinary metal, the renormalized Coulomb-potential reads $V(r) = V_0(r)\e{-r/\lambda}$, where $\lambda \equiv [N(E_\text{F})]^{-1/2}$ is the Thomas-Fermi screening length and $N(E_\text{F})$ is the density of states (DOS) at Fermi level. Since pure graphene has zero DOS at Fermi level, one 
might quite reasonably  suspect that the screening of charge vanishes, and it might seem paradoxical that Coulomb interactions can be neglected. The resolution to this is found by realizing that one may disregard the Coulomb interaction if it is weak compared to the kinetic energy in the problem. Due to the linear dispersion, the kinetic energy is governed by the Fermi velocity $v_\text{F}$ which formally diverges near Fermi level. 
The divergence is logarithmic and precisely due to the Coulomb interaction \cite{gonzales1999,kane2004}. The limiting velocity in graphene 
(due to \eg Umklapp processes) is nevertheless of order $\mathcal{O}(10^6)$ m/s, see e.g. Ref. \onlinecite{novoselov_nature}. This is roughly 100 times larger than in a normal metal, and it is thus safe to neglect the Coulomb interaction compared to the kinetic energy in graphene. In graphene, the Coulomb interaction self-destructs. 
\par
In this work, we will study in detail how an anisotropic order parameter induced in graphene will affect quantum transport in a N/S and N/I/S junction, extending the result of Ref.~\onlinecite{linderPRL07}. In equivalent metallic junctions, it is well-known \cite{hu} that the zero bias conductance peak (ZBCP) is an experimental signature of anisotropic superconductivity in clean superconductors with nodes in the gap. This is a consequence of bound surface states with zero energy at the interface that form due to a constructive phase-interference between electron-like and hole-like transmissions into the superconductor \cite{tanaka}. In graphene junctions with superconductors, as we shall see, a new phenomenology comes into play with regard to the scattering processes that take place at the N/S interface. It is therefore desirable to clarify how anisotropic superconductivity is manifested in the conductance spectra of such a junction, and in particular if the same condition for formation of a ZBCP holds for graphene junctions as well. As first shown in Ref.~\onlinecite{linderPRL07}, we will demonstrate that in N/I/S graphene junctions, novel conductance-oscillations as a function of bias voltage are present both for $s$-wave and $d$-wave symmetry of 
the superconducting condensate due to the presence of low-energy `relativistic' nodal fermions on the N-side. 
The period of the oscillations decreases with increasing width $w$ of the insulating region, and persists even 
if the Fermi energy in I is strongly shifted. This contrasts sharply with metallic N/I/S junctions, where the 
presence of a potential barrier causes the transmittance of the junction to go to zero with increasing $w$. The 
feature of conductance-oscillations is thus unique to N/I/S junctions with low-energy Dirac-fermion excitations. Moreover, we contrast the N/S or N/I/S conductance spectra for the cases where $s$-wave and $d_{x^2-y^2}$-wave 
superconductor constitutes the S-side. The former has no nodes in the gap and lacks Andreev bound 
states. The latter has line-nodes that always cross the Fermi surface in the gap, and thus features
in addition to Andreev bound states, also nodal relativistic low-energy Dirac fermions. The quantum 
transport properties in a heterostructure of two such widely disparate systems, both featuring a particular 
intriguing emergent low-energy symmetry, is of considerable importance. 
\par
This paper is organized as follows. In Sec. \ref{sec:theory}, we establish the theoretical framework which we shall adopt in our treatment of the N/S graphene junction. The results are given in Sec. \ref{sec:results}, where we in particular treat the role of the barrier strength and doping with respect to how the conductance is influenced by these quantities. In addition, we investigate the role of a possible unconventional pairing symmetry induced in graphene. A discussion of our findings is given in Sec. \ref{sec:discussion}, and we summarize in Sec. \ref{sec:summary}. We will use $\check{...}$ for $4\times4$ matrices, and $\hat{...}$ for $2\times2$ matrices, with boldface notation for three-dimensional row vectors.

\section{Theoretical formulation}\label{sec:theory}
\subsection{General considerations}
\begin{figure}[h!]
\centering
\resizebox{0.5\textwidth}{!}{
\includegraphics{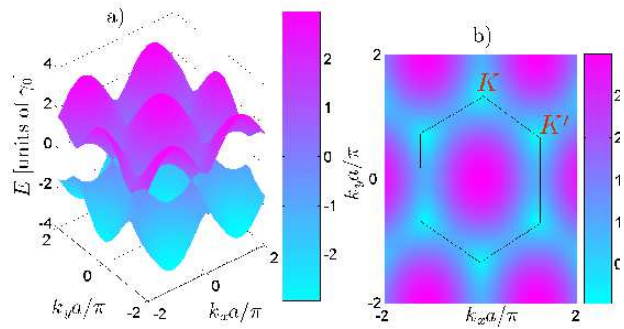}}
\caption{(Color online) a) The energy dispersion for graphene in the Brillouin zone. The upper band is the antibonding $\pi$-orbital, while the lower band is the bonding $\pi$-orbital. It is seen that the bands touch at Fermi level ($E_\text{F}=0$) at six discrete points, which constitutes the effective Fermi surface. b) Contour plot of the dispersion relation, clearly showing the hexagonal structure of the Fermi points. The center of each red drop-like structure represents either $K$ or $K'$.}
\label{fig:dispersion}
\end{figure}

The Brillouin zone of graphene is hexagonal and the energy bands touch the Fermi level at the edges of this zone, amounting to six discrete points. Out of these, only two are inequivalent, which are conventionally dubbed $K$ and $K'$, and referred to as Dirac points. The band dispersion of graphene was first calculated by Wallace \cite{wallace}, and reads
\begin{equation}
E = \pm \gamma_0 \Big[1 + 4\cos\Big(\frac{\sqrt{3}k_xa}{2}\Big)\cos\Big(\frac{k_ya}{2}\Big) + 4\cos^2\Big(\frac{k_ya}{2}\Big)\Big]^{1/2},
\end{equation}
where $\gamma_0 \simeq 2.5$ eV, and the $\pm$ sign refers to the anti-bonding/bonding $\pi$-orbital. The remaining three valence electrons are in hybridized $sp^2$ $\sigma$-bonds. The energy dispersion in the Brillouin zone is plotted in Fig. \ref{fig:dispersion}, which reveals the conical structure of the conduction and valence bands at the six Fermi points. The cosine-like conduction and valence bands are made up by a mixture of the energy bands from the A- and B-sublattices in graphene (Fig. \ref{fig:honeycomb}), which are linear near the Fermi level. This gives rise to the conical energy dispersion at the Dirac points $K$ and $K'$.
\par
\begin{figure}[h!]
\centering
\resizebox{0.38\textwidth}{!}{
\includegraphics{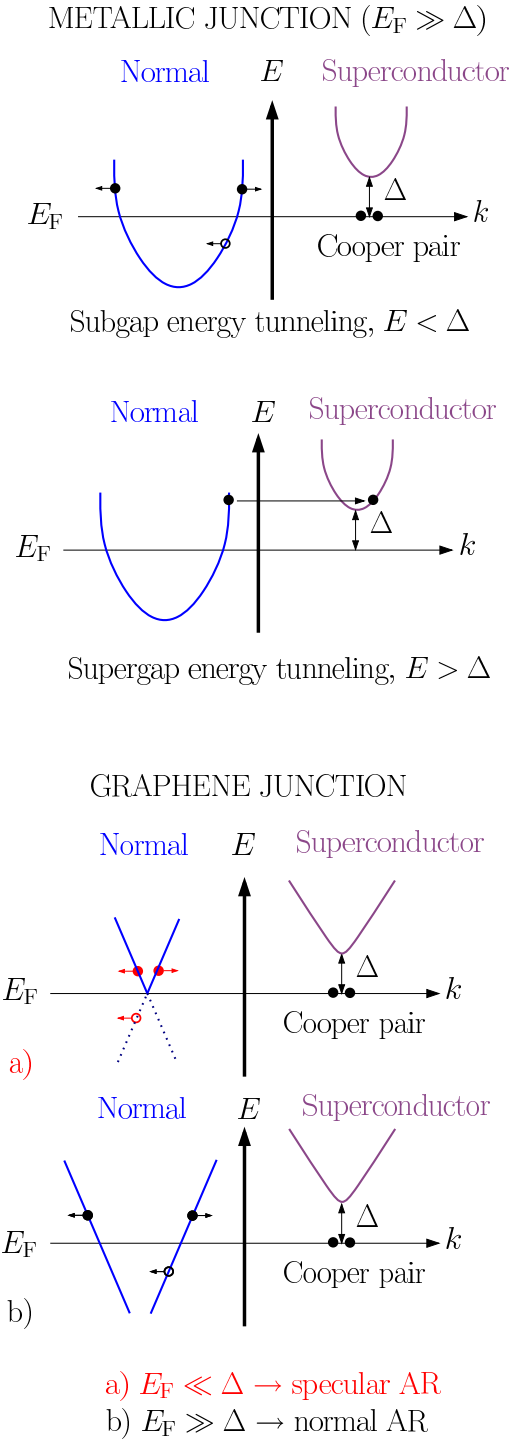}}
\caption{(Color online) Graphical illustration of the different scattering processes that may take place at 
a \textit{i)} metallic and \textit{ii)} graphene N/S junction. While the Andreev reflected hole in case 
\textit{i)} retraces the trajectory of the incoming electron, the hole in case \textit{ii)} may be specularly reflected. This peculiar property is a result of the existence of two bands (conduction and valence) close to 
the Fermi energy in graphene. In the low-energy transport regime, \ie quasiparticle energies $E$ of order $\mathcal{O}(\Delta)$, retroreflection dominates if $E_\text{F} \gg \Delta$, while specular reflection 
dominates if $E_\text{F} \ll \Delta$.}
\label{fig:compare}
\end{figure}

In order to introduce the new phenomenology of scattering processes in N/S graphene junctions, it is instructive to compare it with the metallic N/S junction. This is done in Fig. \ref{fig:compare}. In the metallic case, an incident electron with energy $E < \Delta$ measured from the Fermi energy $E_\text{F}$ can not transmitted into the superconductor since there are no available quasiparticle states. Instead, it is reflected as a hole, represented as a quasiparticle with energy $E$ in the hole-like band, such that the leftover charge $2e$ is transferred into the superconductor as a Cooper pair at Fermi level. The hole has negative mass, energy, wave-vector, and charge compared to the electron which is absent. Strictly speaking, only at $E=0$ are the wave-vectors exactly related through $k_\text{e} = k_\text{h}$ since one in general has
\begin{equation}
k_\text{e}= \sqrt{2m(E_\text{F}+E)},\;\; k_\text{h} = \sqrt{2m(E_\text{F}-E)}.
\end{equation}
At finite energies, the electron-hole coherence will therefore be lost after the hole has propagated a distance 
$L \sim 1/E$. At  energies $E>\Delta$ above the gap, Andreev reflection is severly suppressed since direct tunneling 
into quasiparticle states is now possible.\\
\par
In graphene, a new phenomenology of Andreev reflection is at hand due to the band structure which effectively looks like that of a zero-gap semiconductor (see also Ref.~\onlinecite{beenakkerreview}). Since the conduction and valence bands touch at the Fermi energy $E_\text{F}=0$, one may distinguish between three important cases: \textit{i)} undoped graphene with $E_\text{F}=0$, \textit{ii)} 
doped graphene with $E_\text{F} > 0$, and \textit{iii)} heavily doped graphene with $E_\text{F} \gg 0$. These 
different scenarios are shown in Fig. \ref{fig:compare}.
\par
In undoped graphene, with $E_\text{F} \ll \Delta$, an incident electron with energy $E$ is denied access as a quasiparticle into the superconductor, and physical transport across the junction is thus manifested through 
reflection as a hole. When Andreev reflection takes place, the transmitted Cooper pair is located at the Fermi level of the superconductor. 
Energy conservation then demands that the missing electron in the normal region that is reflected as a hole must be located at $-E$ due to the energy conservation, i.e. in the valence band. This is 
different from normal Andreev reflection, since in that case both the electron and hole belong to the same band (conduction). 
For specular Andreev reflection, however, they belong to \textit{different} bands. The use of the term "specular" 
in order to characterize this type of Andreev reflection originates with the fact that the group velocity $\mathbf{v}_\text{g}$ and momentum $\vk$ have the same sign for a valence band hole, while in contrast $\mathbf{v}_\text{g}$ and $\vk$ have opposite signs for a conduction band hole. To see this, consider first a usual metallic parabolic dispersion $E = \vk^2/2m - E_\text{F}$ for the electrons, such that one readily infers from $\vg = \nabla_\vk E$ 
that $\vg = \vk/m$. Therefore, a hole created at a given energy $E$ will have $\vg = -\vk/m$, since holes have 
opposite group velocities of the electrons for a given wave-vector $\vk$. For normal Andreev reflection, the 
holes are located in the conduction band and therefore satisfy $\vg$ $||$ $-\vk$. \\
\par
In the case of specular Andreev reflection for undoped graphene $(E_\text{F}=0)$, a hole is generated in the valence band. Since in the valence band, the electronic dispersion reads $E = -v_\text{F}|\vk|$, the group velocity of electrons is opposite to their momentum. Conversely, the group velocity for valence holes is parallell to their momentum. This is the mechanism behind specular Andreev reflection. In doped graphene ($E_\text{F} > 0$), the Andreev reflection can be normal or specular, depending on the energy of the incoming electron, as sketched in Fig. \ref{fig:compare}. In heavily doped graphene, ($E_\text{F} \gg \Delta)$, only normal Andreev reflection is present for subgap energies since the distance from Fermi level to the valence band is too large for specular AR to occur. In the regime $E_\text{F} \in$ [0,$\Delta$], one has either normal or specular Andreev reflection, depending on the incident electron energy $E$.\\
\par
\begin{figure}[h!]
\centering
\resizebox{0.40\textwidth}{!}{
\includegraphics{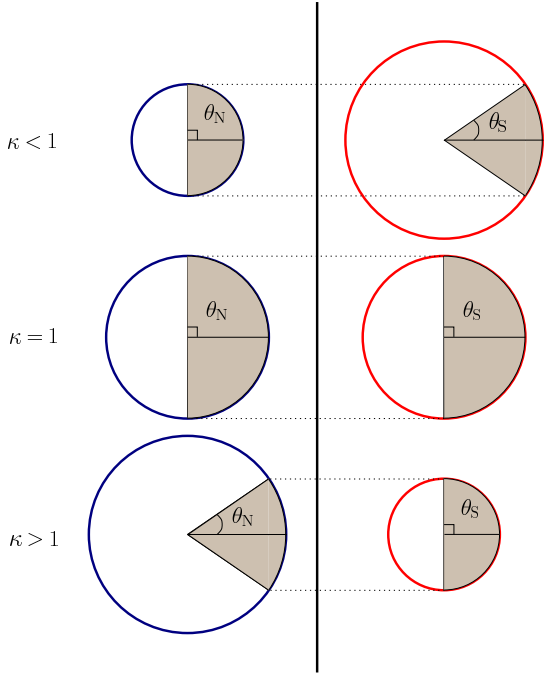}}
\caption{(Color online) The effective impedance caused by FVM illustrated schematically for all possible cases 
of smaller, equal, and larger Fermi velocity in the normal part of the system. Except for the case when the 
Fermi velocities are identical in the normal and superconducting part of the system ($\kappa=1$), part of the 
Fermi surface does not participate in the scattering processes, resulting in a reduction in conductance. For 
instance, when $\kappa<1$, total reflection occurs at angles $\theta_N > \text{asin}(\kappa)$, such that FVM
effectively acts as a source of normal reflection \cite{kashiwaya96}. }
\label{fig:FS}
\end{figure}

We also comment on the effect of Fermi vector mismatch (FVM). Blonder and Tinkham \cite{bt} showed that in a metallic N/S junction, a FVM would act as a source for normal reflection, such that one could effectively account for it simply by choosing a higher value for the barrier strength $Z$. Interestingly, in a ferromagnet/superconductor junction the effect of a FVM could not be reproduced by simply shifting $Z$ to a higher value, as discussed by Zutic and Valls \cite{zutic}. In the absence of an exchange energy, however, the effect of FVM can be thought of as a reduction of the Fermi surface that participates in the scattering processes, as illustrated in Fig. \ref{fig:FS}. One may parametrize the FVM by the parameter $\kappa = k_\text{F}/q_\text{F}$ where $k_\text{F}$ ($q_\text{F}$) is the Fermi momentum in the normal (superconducting) part of the system. In particular, it is seen that for $\kappa > 1$, there is only possible transmission of quasiparticles (although these decay exponentially) up to a critical angle less than 
$\pi/2$ \cite{kashiwaya96}.

Having established the states that participate in the scattering at the interface, we now turn to equations 
that describe these quasiparticle states. 
\subsection{Scattering processes}
Consider the case of zero external magnetic field. The full Bogoliubov-de Gennes (BdG) equation for the 
2D sheet graphene normal/$s$-wave superconductor junction in the $xy$-plane then reads \cite{beenakker, beenakkerreview}
\begin{align}\label{eq:Bdg1}
\begin{pmatrix}
\check{H} - E_\text{F}\check{1} & \Delta_\vk \check{1} \\
\Delta_\vk^\dag \check{1} & E_\text{F}\check{1} - \check{\mathcal{T}}\check{H}\check{\mathcal{T}}^{-1}\\
\end{pmatrix}
\begin{pmatrix}
u\\
v\\
\end{pmatrix}
 = E \begin{pmatrix}
 u\\
 v\\
 \end{pmatrix}, 
\end{align}
where $E$ is the excitation energy, and $\{u,v\}$ denoting the electron-like and hole-like exictations described by the wave-function. Assuming that the superconducting region is located at $x>0$ and neglecting the decay of the order parameter in the vicinity of the interface \cite{bruder}, we may write for the spin-singlet order parameter
\begin{equation}
\Delta_\vk = \Delta(\theta)\e{\i\vartheta}\Theta(x),
\end{equation}
where $\Theta(x)$ is the Heaviside step function, $\vartheta$ is the phase corresponding the globally broken $U(1)$ symmetry in the superconductor, while $\theta = \text{atan}(k_y/k_x)$ is the angle on the Fermi surface in reciprocal space (we have adopted the weak-coupling approximation with $\mathbf{k}$ fixed on the Fermi surface). Note that in contrast to previous work, we allow for the possibility of unconventional superconductivity in the graphene layer since $\Delta_\vk$ now may be anisotropic. We have applied weak-coupling limit, the momentum $\vk$ is fixed on the Fermi surface, such that $\Delta_\vk$ only has an angular dependence. Since we employ a spin-singlet even parity order parameter, the condition $\Delta(\theta) = \Delta(\pi+\theta)$ must be fulfilled. The single-particle Hamiltonian is given by
\begin{align}
\check{H} = \begin{pmatrix}
\hat{H}_+ & 0\\
0 & \hat{H}_- \\
\end{pmatrix},\;\;
\hat{H}_\pm = -\i v_\text{F} (\hat{\sigma}_x\partial_x \pm \hat{\sigma}_y\partial_y).
\end{align}
Here, $v_\text{F}$ is the energy-independent Fermi velocity for graphene, while $\hat{\sigma}_i$ denotes the Pauli matrices. For later use, we also define the Pauli matrix vector $\hat{\boldsymbol{\sigma}} = (\hat{\sigma}_x, \hat{\sigma}_y, \hat{\sigma}_z)$. These Pauli matrices operate on the sublattice space of the honeycomb structure, corresponding to the A and B atoms, while the $\pm$ sign refers to the two so-called \textit{valleys} of $K$ and $K'$ in the Brillouin zone. The Dirac points earn their sobriquet as valleys from the geometrical resemblance of the band dispersion to the aforementioned. The spin indices may be suppressed since the Hamiltonian is time-reversal invariant. In addition to the spin degeneracy, there is also a valley degeneracy, which effectively allows one to consider either the one of the $\hat{H}_\pm$ set. Therefore, the $8 \time 8$ matrix BdG-equation Eq. (\ref{eq:Bdg1}) reduces to a 
$4 \times 4$ matrix BdG-equation, namely
\begin{align}
\begin{pmatrix}
\hat{H}_\pm - E_\text{F}\hat{1} & \Delta_\vk \hat{1} \\
\Delta_\vk^\dag \hat{1} & E_\text{F}\hat{1} - \hat{H}_\pm\\
\end{pmatrix}
\begin{pmatrix}
u\\
v\\
\end{pmatrix}
 = E \begin{pmatrix}
 u\\
 v\\
 \end{pmatrix}, 
\end{align}
where have explicitly used that $\check{\mathcal{T}}\check{H} =\check{H}\check{\mathcal{T}}$. Let us then consider $\hat{H}_+$, such that one may write
\begin{align}\label{eq:Bdg2}
\begin{pmatrix}
\vp\cdot\pauli - E_\text{F}\hat{1} & \Delta_\vk \hat{1} \\
\Delta_\vk^\dag \hat{1} & E_\text{F}\hat{1} - \vp\cdot\pauli\\
\end{pmatrix}
\begin{pmatrix}
u\\
v\\
\end{pmatrix}
 = E \begin{pmatrix}
 u\\
 v\\
 \end{pmatrix}.
\end{align}
In the above Hamiltonian, we have only included diagonal terms in the gap matrix, \ie $\hat{\Delta}_\vk = \Delta_\vk\hat{1}$. This corresponds to exclusively intraband-pairing on each of the sublattices A and B. In recent work by Black-Schaffer and Doniach \cite{blackschaffer}, it was shown that by postulating interband spin-singlet hopping between the sublattices, one could achieve dominant $d$-wave pairing in intrinsic graphene. While an onsite attractive potential is sufficient to achieve $s$-wave pairing, leading to a diagonal gap matrix, nearest-neighbor interactions couples the two sublattices and should yield off-diagonal elements in the gap matrix. In this work, we restrict ourselves to anisotropic superconducting pairing with diagonal elements in the gap matrix, although one would have to take into account off-diagonal elements as well for a completely general treatment. We comment more on this later.
\par
\begin{figure}[h!]
\centering
\resizebox{0.475\textwidth}{!}{
\includegraphics{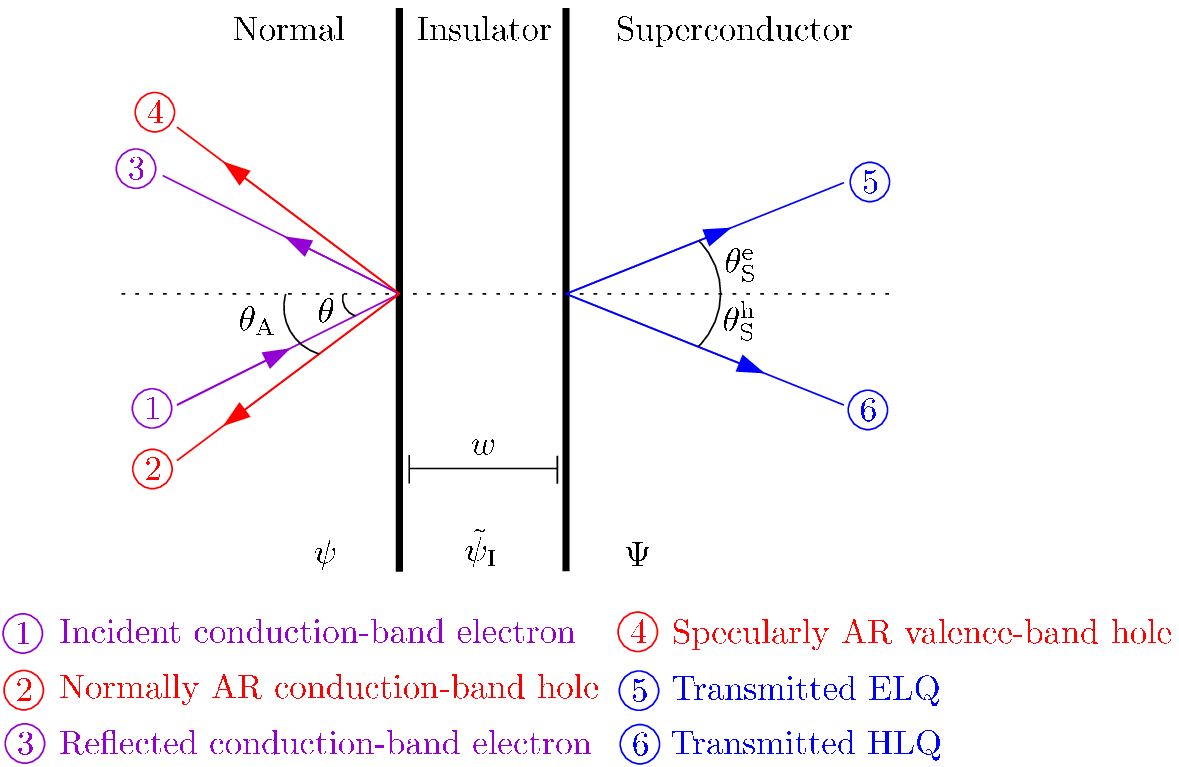}}
\caption{(Color online) The scattering processes taking place at an N/S or N/I/S graphene junction. In the former case, the insulating region is completely absent, and only the six depicted processes take place. Note that only normal Andreev reflection \textit{or} specular Andreev reflection takes place at any given energy $E$, never both. For an N/I/S graphene junction, there are transmitted and reflected electrons and holes in the insulating region corresponding to $\tilde{\psi}_\text{I}$, not shown in the above figure. }
\label{fig:scattering}
\end{figure}

Consider an incident electron from the normal side of the junction $(x<0)$ with energy $E$. For positive excitation energies $E>0$, the eigenvectors and corresponding momentum of the particles read 
\begin{align}
\psi^\text{e}_+ &= [1, \e{\i\theta}, 0,0]^\text{T}\e{\i p^\text{e}\cos\theta x},\; p^\text{e} = (E +  E_\text{F})/v_\text{F},
\end{align}
for a right-moving electron at angle of incidence $\theta$ (see Fig. \ref{fig:scattering}, while a left-moving 
electron is described by the substitution $\theta\to\pi-\theta$. If Andreev-reflection takes place, a left-moving 
hole is generated with an energy $E$, angle of reflection $\theta_\text{A}$, and corresponding wave-function
\begin{align}
\psi^\text{h}_- = [0,0,1,\e{-\i\theta_\text{A}}]^\text{T}\e{-\i p^\text{h}\cos\theta_\text{A} x},\; p^\text{h} = (E - E_\text{F})/v_\text{F},
\end{align}
where the superscript e (h) denotes an electron-like (hole-like) excitation. Since translational invariance in 
the $\hat{\mathbf{y}}$-direction holds, the corresponding component of momentum is conserved. This condition 
allows for determination of the Andreev-reflection angle $\theta_\text{A}$ through $p^\text{h}\sin\theta_\text{A} = p^\text{e}\sin\theta.$ From this equation, one infers that there is no Andreev-reflection ($\theta_\text{A} = \pm\pi/2$), and consequently no subgap conductance, for angles of incidence above the critical angle 
\begin{equation}
\theta_\text{c} = \text{asin}\Big( \frac{|E-E_\text{F}|}{E+E_\text{F}} \Big).
\end{equation}
On the superconducting side of the system ($x>w$), the possible wavefunctions for transmission of a right-moving quasiparticle with a given excitation energy $E>0$ reads
\begin{align}\label{eq:psisuper}
\Psi^\text{e}_+ &= \Big(u(\theta^+), u(\theta^+)\e{\i\theta^+},v(\theta^+)\e{-\i\phi^+},v(\theta^+)\e{\i(\theta^+-\phi^+)}\Big)^\text{T}\notag\\
&\times\e{\i q^\text{e}\cos\theta^+ x},\; q^\text{e} = (E'_\text{F} + \sqrt{E^2-|\Delta(\theta^+)|^2})/v_\text{F},\notag\\
\Psi^\text{h}_- &= \Big(v(\theta^-), v(\theta^-)\e{\i\theta^-},u(\theta^-)\e{-\i\phi^-},u(\theta^-)\e{\i(\theta^--\phi^-)}\Big)^\text{T}\notag\\
&\times\e{\i q^\text{h}\cos\theta^- x},\; q^\text{h} = (E'_\text{F} - \sqrt{E^2-|\Delta(\theta^-)|^2})/v_\text{F}.
\end{align}
The coherence factors are, as usual, given by \cite{sudbo}
\begin{align}
u(\theta) &= \sqrt{\frac{1}{2}\Big(1 + \frac{\sqrt{E^2-|\Delta(\theta)|^2}}{E}\Big)},\notag\\
v(\theta) &= \sqrt{\frac{1}{2}\Big(1 - \frac{\sqrt{E^2-|\Delta(\theta)|^2}}{E}\Big)}.
\end{align}
Above, we have defined $\theta^+ = \theta_\text{S}^\text{e}$, $\theta^- = \pi-\theta_\text{S}^\text{h}$, and $\e{\i\phi^\pm} = \e{\i\vartheta}\Delta(\theta^\pm)/|\Delta(\theta^\pm)|$.
The transmission angles $\theta^\text{(i)}_\text{S}$ for the electron-like (ELQ) and hole-like (HLQ) quasiparticles 
are given by $q^\text{(i)}\sin\theta^\text{(i)}_\text{S} = p^\text{e} \sin\theta$, i$=$e,h. Note that for subgap energies $E<\Delta$, there is a small imaginary contribution to the wavevector, which leads to exponentional damping of the wavefunctions inside the superconductor. The physical reason for this is that there can be no transmission of quasiparticles into the superconductor for subgap energies. Note that for all wavefunctions listed in the equations above, we have for clarity not included a common phase factor $\e{\i k_yy}$ which corresponds to the conserved momentum in the $\hat{\mathbf{y}}$-direction. A possible Fermi vector mismatch (FVM) between the normal and superconducting region is accounted for by allowing for $E'_\text{F} \neq E_\text{F}$. The case $E'_\text{F} \gg \Delta$ corresponds to a heavily doped superconducting region. It is also
straight-forward to obtain the eigenfunctions for the case when the gap matrix consists of off-diagonal elements, as opposed to the gap matrix treated here with diagonal elements. In particular, for
\begin{equation}
\hat{\Delta}_\vk = \begin{pmatrix}
0 & \Delta(\theta)\e{\i\nu} \\
\Delta(\theta)\e{\i\nu} & 0\\
\end{pmatrix},
\end{equation}
the eigenfunctions may be obtained from Eq. (\ref{eq:psisuper}) simply by switching the phase-factors as follows:
\begin{align}\label{eq:psisuper}
\Psi^\text{e}_+ &= \Big(u(\theta^+), u(\theta^+)\e{\i\theta^+},v(\theta^+)\e{\i(\theta^+-\phi^+)},v(\theta^+)\e{-\i\phi^+}\Big)^\text{T}\notag\\
&\times\e{\i q^\text{e}\cos\theta^+ x},\; q^\text{e} = (E'_\text{F} + \sqrt{E^2-|\Delta(\theta^+)|^2})/v_\text{F},\notag\\
\Psi^\text{h}_- &= \Big(v(\theta^-), v(\theta^-)\e{\i\theta^-},u(\theta^-)\e{\i(\theta^--\phi^-)},u(\theta^-)\e{-\i\phi^-}\Big)^\text{T}\notag\\
&\times\e{\i q^\text{h}\cos\theta^- x},\; q^\text{h} = (E'_\text{F} - \sqrt{E^2-|\Delta(\theta^-)|^2})/v_\text{F}.
\end{align}
\par At this stage, it is appropriate to insert the restriction which will be used throughout the rest of this paper, namely $\Delta \ll E_\text{F}'$. Since we are using a mean-field approach to describe the superconducting part of the Hamiltonian, it is implicitly understood that phase-fluctuations of the order parameter must be small \footnote{It is sufficient to demand that the phase-fluctuations must be small, since it was shown by Kleinert \cite{kleinert} that for any system exhibiting a second-order phase transition with a spontaneously broken O($N$) $(N \geq 2)$ symmetry at low temperatures, phase-fluctuations destroy order before amplitude-fluctations become important.}. For this criteria to be fulfilled, the superconducting coherence length $\xi$ must be large compared to some characteristic length scale of the system \cite{kleinert}. Following Ref.~\onlinecite{kleinert}, the critical temperature $T_\text{K}$ at which long-range phase-fluctuations of the order parameter destroys the ordering when approaching the critical temperature $T_\text{c}$ from below, is given by
\begin{equation}
T_\text{K} = T_\text{c}^\text{MF}(1 - |\tau|),
\end{equation} 
where $|\tau| \sim \xi^D$, $D$ is the dimensionality of the system, and $T_\text{c}^\text{MF}$ is the critical temperature predicted by mean-field theory. (For an extensive treatment of the effect of phase fluctuations
in extreme type-II superconductors, see Ref. \onlinecite{phase_fluctuations}.)
Thus, only for $|\tau| \ll 1$  mean-field theory is a viable option for describing superconductivity in the system, corresponding to a large coherence length $\xi$. Notice that the Ginzburg temperature $T_\text{G}$, which describes the regime where amplitude-fluctuations of the order parameter become important $(T>T_\text{G})$, satisfies $T_\text{c}^\text{MF} > T_\text{G} > T_\text{K}$. A natural choice of characteristic length scale for the system in the normal state is obviously the Fermi wavelength $\lambda_\text{F}' = 2\pi v_\text{F}'/E_\text{F}'$, such that the criteria for validity of mean-field theory reads $\xi/\lambda_\text{F}' \gg 1$, or equivalently, $E_\text{F}' \gg \Delta$. 
 \par
 The relevant scattering processes at the N/S graphene interface are shown in Fig. \ref{fig:scattering}, in the two cases of zero barrier and an insulating interface of width $w$. In the former case, the boundary conditions dictate that $\psi|_{x=0} = \Psi|_{x=0}$, where
\begin{align}
\psi &= \psi^\text{e}_+ + r\psi^\text{e}_- + r_\text{A}\psi^\text{h}_-,\notag\\
\Psi &=  t^\text{e}\Psi^\text{e}_+ + t^\text{h}\Psi^\text{h}_-,
\end{align}
while in the latter case, one must match the wavefunctions at both interfaces:
\begin{align}\label{eq:nisboundary}
\psi|_{x=0} = \tilde{\psi}_ \text{I}|_{x=0}, \;\tilde{\psi}_\text{I}|_{x=w} = \Psi_\text{S}|_{x=w}
\end{align}
where we have defined the wavefunction in the insulating region
\begin{align}
\tilde{\psi}_\text{I} = \tilde{t}_1\tilde{\psi}^\text{e}_+ + \tilde{t}_2\tilde{\psi}^\text{e}_- + \tilde{t}_3\tilde{\psi}^\text{h}_+ + \tilde{t}_4\tilde{\psi}^\text{h}_-.
\end{align}
The wavefunctions $\tilde{\psi}$ differ from $\psi$ in that the Fermi energy is greatly shifted by means of \eg an external potential, such that $E_\text{F} \to E_\text{F}-V_0$ where $V_0$ models the potential barrier (equivalent to the role of $Z$ in Ref.~\onlinecite{btk}). Also, note that the trajectories of the quasiparticles in the insulating region, defined by the angles $\tilde{\theta}$ and $\tilde{\theta}_\text{A}$, differ by the same substitution, meaning
\begin{align}
\sin\tilde{\theta}/\sin\theta = (E+E_\text{F})/(E+E_\text{F}-V_0),\notag\\
\sin\tilde{\theta}_\text{A}/\sin\theta = (E+E_\text{F})/(E-E_\text{F}+V_0).\notag\\
\end{align}
Finally, note that the subscript $\pm$ on the wavefunctions in the normal region indicates the direction of their group velocity, which \textit{in general} is different from the direction of momentum, as discussed previously. Consequently, although the Andreev-reflected hole wavefunction carries a subscript "$-$" above, one should keep in mind that for normal Andreev reflection, the direction of momentum is opposite to the group velocity for the hole.

\par
Before we go on to presenting results, we make one conceptual remark. 
The proximity effect means that an otherwise normal system becomes superconducting by virtue of having the superconducting wavefunction from a nearby superconductor leak into the normal system, thus making it superconducting in some region. This is a result of a boundary condition imposed on the normal system from the proximate host superconductor. The resulting wavefunction in the proximity region of the normal system is then a BCS type wavefunction. Such a wavefunction unquestionably describes a system with a gapped Fermi surface (possibly with nodes on the Fermi-surface). It matters not by what microscopic mechanism such a state was established, as long as it is there. The effective interaction giving rise to proximity induced superconductivity in graphene close to the surface in contact with an intrinsically superconducting host system, is obtained by considering the complete superconductor-graphene system and integrating out the electrons on the superconducting side. The electrons in graphene then experience an effective attractive interaction $\lambda_\text{eff}$ giving rise to a gap by virtue of hopping into and out of the superconducting side. We thus have, by such tunneling processes, $\lambda_\text{eff} \neq 0$ even if $\lambda =0$. Here, $\lambda$ is the electron-electron coupling constant giving rise to superconductivity in graphene {\it per se}.
Since graphene intrinsically is a normal system, and is well approximated by non-interacting electrons, this
coupling constant vanishes, $\lambda=0$.  The relationship between the gap in the normal region $\Delta$ and $\langle ff \rangle$ is thus 
$\Delta = \lambda_\text{eff} \langle ff \rangle$, and this gives a nonzero gap in the vicinity of the proximate host superconductor. Here, $f$ are fermion annhilation operators, and $\langle ff \rangle$ thus represents the pair-amplitude induced in the normal graphene region.
The proximity-induced gap vanishes rapidly as one goes away from the surface and into the bulk of graphene, since 
$\lambda_\text{eff}$ vanishes rapidly as we move away from the proximate host superconductor. 
It would in principle be {\it incorrect} to assert that in the proximity-region of graphene, we could have a non-zero anomalous 
Green's function $\langle ff \rangle$,  but no gap $\Delta$, by using a self-consistency relation of
the type $\Delta = \lambda \langle ff \rangle$ with $\lambda=0$ \cite{RevModPhys_2005}. Such a self-consistency relation 
does not exist in a normal system which does not superconduct by itself.  However, in a situation where the intrinsic 
$\lambda =0$, it could well turn out to be the case that $\lambda_\text{eff}$ is small, leading to a gap which is very 
small \cite{bruder}. In our paper, we have a thin film graphene system with a bulk superconductor in contact with the film, 
deposited on top of the film. If the thickness of the graphene-film is smaller than the coherence length of the bulk superconductor, 
one obtains a proximity-induced superconductiving gap throughout the film. As we shall see, our results are quite sensitive 
to the presence of even a small induced gap in graphene.

\section{Conductance spectra}\label{sec:results}
In what follows, we describe how the conductance spectra of a N/S and N/I/S graphene junction may be obtained. According to the BTK formalism \cite{btk}, the normalized conductance is given by
\begin{align}\label{eq:conductance}
G(eV) = \frac{1}{G_\text{N}} \int^{\pi/2}_{-\pi/2} &\text{d}\theta\cos\theta \Big(1 -|r(eV,\theta)|^2  \notag\\
&+ \frac{\cos\theta_\text{A}}{\cos\theta}|r_\text{A}(-eV,\theta)|^2\Big),
\end{align}
where $r$ and $r_\text{A}$ are the reflection coefficients for normal and Andreev reflection \footnote{Note that in Ref.~\onlinecite{linderPRL07}, a factor $|p_h/p_e|$ was included as a prefactor of $|r_\text{A}(-eV,\theta)|^2$. While such a factor is present in the metallic case, it is absent for graphene junctions. Nevertheless, this does not affect the results of Ref.~\onlinecite{linderPRL07} in any manner since the case $E_F\gg(\Delta,\varepsilon)$ was considered there, implying $|p_h/p_e|\approx1$.}, respectively, while $G_\text{N}$ is a renormalization constant corresponding to the N/N metallic conductance \cite{kashiwaya96},
\begin{equation}
G_\text{N} = \int^{\pi/2}_{-\pi/2} \text{d}\theta \cos\theta\frac{4\cos^2\theta}{4\cos^2\theta + Z^2}.
\end{equation}
In this case, we have zero intrinsic barrier such that $Z=0$.
We will apply the usual approximation $|r_\text{A}(-eV,\theta)| = |r_\text{A}(eV,\theta)|$, which may be shown to hold for a quite general parameter regime. For perfect normal reflection ($|r|^2=1$), there is no conductance, while for perfect Andreev reflection ($|r_\text{A}|^2=1$), the conductance is doubled compared to the N/N case. In order to obtain these coefficients, we make use of the boundary conditions described in the previous section. The analytical solution and behaviour of the conductance differs in the N/S and N/I/S case, and we proceed with a separate treatment of these scenarios.

\subsection{N/S junction}
Solving the boundary conditions for the wavefunctions at the interface leads to the analytical expressions for the reflection coefficients:
\begin{align}\label{eq:coeff}
&r = \frac{2\cos\theta[\zeta_+v(\theta^+)v(\theta^-)\e{\i(\phi^--\phi^+)} - \zeta_-u(\theta^+)u(\theta^-)]}{v(\theta^+)v(\theta^-)\e{\i(\phi_--\phi_+)}Y_- - u(\theta^+)u(\theta^-)Y_+} - 1,\notag\\
&r_\text{A} = \frac{2\e{-\i\phi^+}\cos\theta[\zeta_+u(\theta^-)v(\theta^+) - \zeta_-u(\theta^-)v(\theta^+)]}{v(\theta^+)v(\theta^-)\e{\i(\phi_--\phi_+)}Y_- - u(\theta^+)u(\theta^-)Y_+},
\end{align}
where we have defined the auxiliary quantities $\zeta_\pm = \e{\i\theta^\pm}  - \e{-\i\theta_\text{A}}$ and $Y_\pm = \zeta_\mp(\e{\i\theta^\pm} + \e{-\i\theta}).$ The interplay between the different phases felt by the ELQ and HLQ in the superconductor in the case of an anisotropic order parameter enters above through the factor $\e{\i(\phi^--\phi^+)}$. It remains, however, to be clarified how this interplay manifests itself in the tunneling conductance. Before investigating this in more detail, let us briefly consider the isotropic $s$-wave case first, \ie $\Delta(\theta) = \Delta$, such that $\e{\i(\phi^--\phi^+)} = 1$.
\subsubsection{Conventional $s$-wave pairing}
\begin{figure}[h!]
\centering
\resizebox{0.475\textwidth}{!}{
\includegraphics{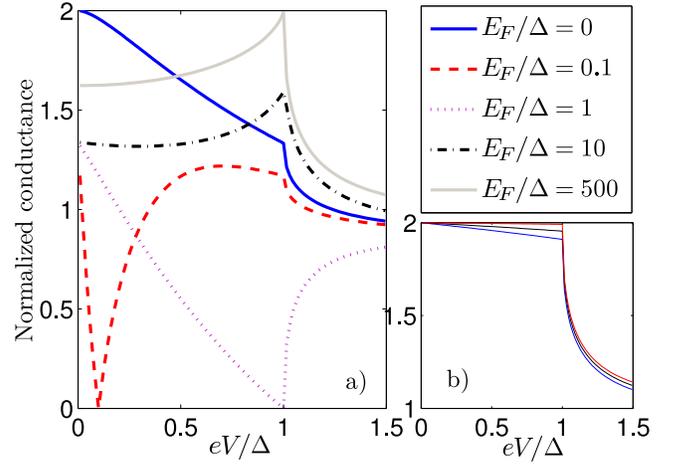}}
\caption{(Color online) Conductance spectra for graphene in a) with $E_\text{F}'/\Delta = 10^3$, and in b) with $E_\text{F}' = E_\text{F}$. The spectra in a) are identical to the result of Ref.~\onlinecite{beenakker}. In b), the subgap conductance is always close to $2G_\text{N}$, but becomes more constant for increasing $E_\text{F}$ since $\theta_\text{c}\to\pi/2$. We have plotted the ratios $\{50, 100, 1000\}$ of $E_\text{F}/\Delta$ in b), from bottom to top.}
\label{fig:swave}
\end{figure}

For conventional superconducting pairing, Eq. (\ref{eq:coeff}) reduces to
\begin{align}\label{eq:coeffswave}
r &= \frac{2\cos\theta(\zeta_+v^2 - \zeta_-u^2)}{v^2Y_- - u^2Y_+} - 1,\notag\\
r_\text{A} &= \frac{2\cos\theta uv(\e{\i\theta^+}-\e{\i\theta^-})}{v^2Y_- - u^2Y_+}.
\end{align}
This case was first studied by Beenakker \cite{beenakker}. For consistency and completeness, we reproduce the results of Ref. ~\onlinecite{beenakker} [see Fig. \ref{fig:swave}a)]. We point out that Eq. (\ref{eq:coeffswave}) are valid for any parameter range, and not restricted to the heavily doped case treated in Ref. ~\onlinecite{beenakker}. To illustrate the difference, we consider the regime $E_\text{F}' = E_\text{F}$ shown in Fig. \ref{fig:swave}b). In this case, the standard situation of perfect Andreev reflection for subgap energies is recovered, with a sharp drop at the gap edge corresponding to the onset of quasiparticle transmittance into the superconductor.

\subsubsection{Anisotropic $d$-wave pairing}
\begin{figure}[h!]
\centering
\resizebox{0.3\textwidth}{!}{
\includegraphics{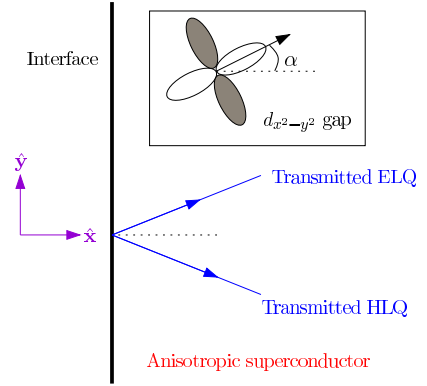}}
\caption{(Color online) Sketch of 2D N/S graphene junction with an anisotropic superconductor. The orientation of the gap in $\vk$-space is modelled by the angle $\alpha$. }
\label{fig:dwaveorientation}
\end{figure}

To treat an unconventional superconducting order parameter, we must revert to the general expressions in Eq. (\ref{eq:coeff}). In order to account for the effect of an anisotropic gap, we choose the $d_{x^2-y^2}$-gap from Tab. \ref{tab:symmetry}, which in the weak-coupling approximation ($|\vk| = k_\text{F}$) reads $\Delta(\theta) = \Delta\cos(2\theta - 2\alpha)$. Here, $\alpha$ models the relative orientation of the gap in $\vk$-space with respect to the interface normal as illustrated in Fig. \ref{fig:dwaveorientation}.

\begin{figure}[h!]
\centering
\resizebox{0.475\textwidth}{!}{
\includegraphics{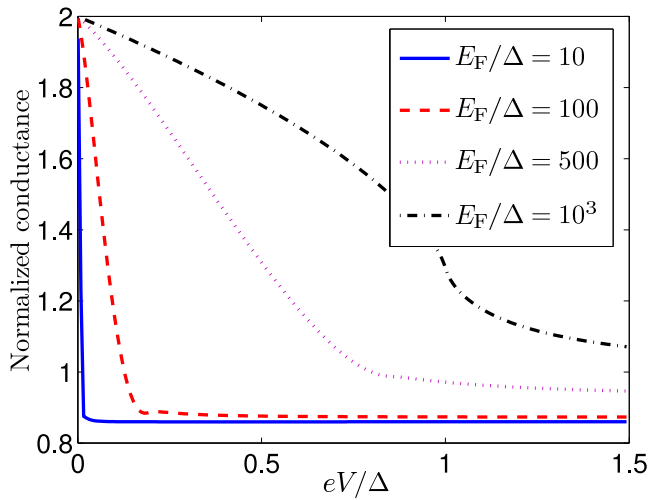}}
\caption{(Color online) Conductance spectra for doped graphene in with $E_\text{F}' = 10^4\Delta$ for a $d$-wave order parameter with orientation angle $\alpha=\pi/4$. A ZBCP is present for large FVM, and becomes unobservable narrow for $E_\text{F}/\Delta < 10$. For $\alpha=0$, the conductance spectra are essentially identical to those in Fig. \ref{fig:swave}.}
\label{fig:dwave}
\end{figure}

We now proceed to investigate how the conductance spectra of a N/S graphene junction change when going from a $s$-wave to a $d$-wave order parameter in the superconducting part of the system. Consider Fig. \ref{fig:dwave} for the case of heavily doped graphene, where the orientation of the gap is such that the condition for perfect formation of zero energy states in a metallic N/S junction is fulfilled, \ie $\Delta(\theta) = -\Delta(\pi-\theta)$. As shown by Tanaka and Kashiwaya \cite{tanaka}, this gives rise to a quasiparticle interference between the ELQ and HLQ since they feel different phases of the pairing potentials due to their different trajectories of transmittance into the superconductor. This results in a bound surface states with zero energy \cite{hu} close to the N/S interface. For the N/S graphene junction studied here, the explicit barrier potential is zero, while FVM effectively acts as as source of normal reflection. From Fig. \ref{fig:dwave}, one may infer that a peak at zero bias is present in the presence of FVM, although the ZBCP does not increase in magnitude with increasing FVM. We will later study how the presence of an intrinsic barrier in the form of a thin, insulating region separating the normal and superconducting part affects the ZBCP.

\begin{figure}[h!]
\centering
\resizebox{0.475\textwidth}{!}{
\includegraphics{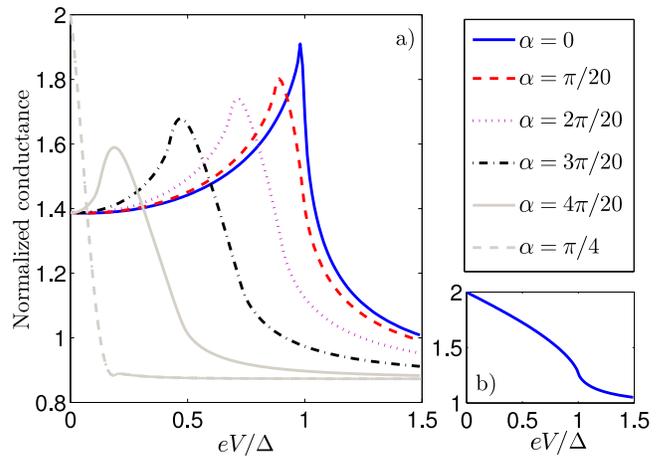}}
\caption{(Color online) Conductance spectra for a) doped graphene in with $E_\text{F}' = 10^4\Delta$ for a $d$-wave order parameter and b) undoped graphene with $E_\text{F} = E_\text{F}'$. In both cases, we have set $E_\text{F}/\Delta = 100$ and investigate how the conductance spectra evolves upon rotating $\alpha$ from $0$ to $\pi/4$ in steps of $\pi/20$. In a), it is seen that the peak of the conductance shifts from $eV=\Delta$ to progressively lower values as $\alpha$ increases. This is in contrast to metallic N/S junctions. In b), we plot the conductance for $\alpha=\pi/4$. We find virtually no difference between various orientations of the gap in this case, and the spectra are quite similar to the metallic N/S case with zero barrier.}
\label{fig:dwave2}
\end{figure}

Next, we plot the conductance spectra for doped graphene to see how they evolve upon a rotation of the gap. The behaviour is quite distinct from that encountered in a N/S metallic junction. From Fig. \ref{fig:dwave2}a), we see that the peak of the conductance shifts from $eV=\Delta$ to progressively lower values as $\alpha$ increases from $0$ to $\pi/4$. In this respect, the conductance spectra actually mimicks a lower value of the gap than what is the case, if one were to infer the gap magnitude from the position of the singularity in the spectra. As of such, for a given FVM, determining the magnitude of the gap by the usual method of locating the characteristic feature in the conductance spectra is not as straight-forwards in N/S graphene junctions as in the metallic case. Indeed, multiple measurements with several different interface orientations would in general be required to obtain the correct value of the gap. This should be a unambigously observable feature in experiments, and provides a direct way of testing our theory. Finally, we consider graphene in Fig. \ref{fig:dwave2}b) with $E_\text{F} = E_\text{F}'$ for a $d$-wave order parameter. Upon varying $\alpha$ from 0 to $\pi/4$, there is now little distinction between different angles of orientation. The conductance spectra are in this case very resemblant to the metallic N/S case for zero barrier \cite{tanaka}.

\subsection{N/I/S junction}
We now consider the conductance of an N/I/S graphene junction, where I denotes an "insulating" (see introduction) region modelled by a very large energy potential for the quasiparticles. Solving the boundary conditions introduced in Sec. \ref{sec:theory}, we obtain analytical expression for $r$ and $r_\text{A}$, which is all that is required in order to calculate the conductance. However, these expressions are very large and the reader may consult Appendix A for their explicit form. In the following, we will not work exclusively in the thin barrier-limit $d\to0, V_0\to\infty$ as in Ref.~\onlinecite{sengupta}. Some aspects of including an insulation region of arbitrary width and strength were very recently discussed in Ref.~\onlinecite{sengupta2}, albeit only in the case of isotropic $s$-wave pairing. We now treat the two cases of $s$-wave and $d$-wave pairing separately.

\subsubsection{Conventional $s$-wave pairing}
\begin{figure}[h!]
\centering
\resizebox{0.475\textwidth}{!}{
\includegraphics{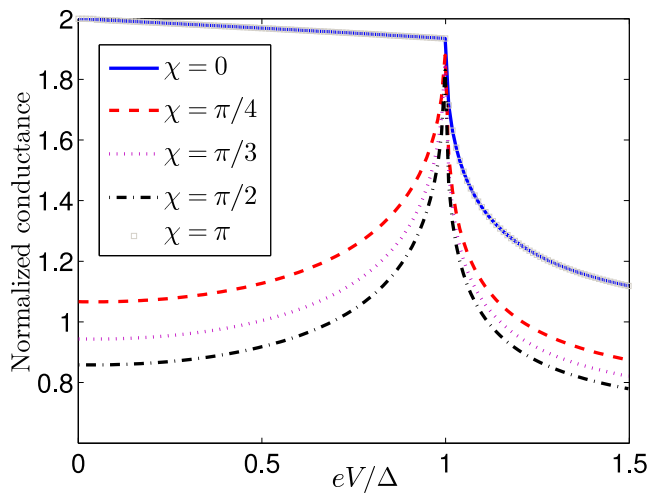}}
\caption{(Color online) Conductance spectra for an N/I/S graphene junction with $E_\text{F}'/\Delta=E_\text{F}/\Delta = 100$, using $s$-wave pairing. We reproduce the same results as Ref.~\onlinecite{sengupta} with a $\pi$-periodicity in the parameter $\chi$. However, we obtain a phase shift of $\pi/2$ in $\chi$ compared to their results. We believe that this difference pertains to a minor sign error in the wavefunctions used in Ref.~\onlinecite{sengupta}.}
\label{fig:nisswave}
\end{figure}

This case was first studied by Bhattacharjee and Sengupta \cite{sengupta}. To quantify the parameters in the insulating region, we will measure the width $w$ of region I in units of $\lambda_\text{F}$ and the potential barrier $V_0$ in units of $E_\text{F}$. First, we briefly show that we are able to reproduce the qualitative findings of Ref.~\onlinecite{sengupta}. As shown in Appendix A, it is convenient to introduce the parameter $\chi = V_0w/v_\text{F}$ in the thin-barrier limit. In this case, the reflection coefficients $r$ and $r_\text{A}$ exhibit an interesting oscillating behaviour as a function of $\chi$. To see this, consider Fig. \ref{fig:nisswave} where we have plotted the voltage dependence of the normalized conductance for several values of $\chi$. For $\chi=0$, we reproduce the result of Fig. \ref{fig:swave}b). This is reasonable since the conductance of an N/I/S junction with $\chi=0$, \ie zero width, should be the same as an N/S junction. The $\pi$-periodicity is reflected in that the curves for $\chi=0$ and $\chi=\pi$ are identical. For $\chi\neq n\pi$, $n=0,1,2,..$, there is a source of normal reflection at the interfaces due to the insulating region, and consequently the subgap conductance is reduced from its ballistic value $2G_\text{N}$. Even in the presence of a FVM, $E_\text{F}' \neq E_\text{F}$, the spectra of Fig. \ref{fig:nisswave} retain their $\pi$-periodicity. However, the FVM acts as a source of normal reflection such that one does not have nearly perfect Andreev reflection at subgap energies. 
\par
Our results differ slightly from those reported in Ref.~\onlinecite{sengupta}. Although we obtain qualitatively exactly the same dependence on $\chi$ of the conductance, it is seen by comparing our Fig. \ref{fig:nisswave} with Fig. 1 of Ref.~\onlinecite{sengupta} that our curves are phase shifted by $\pi/2$ in $\chi$ in comparison. As a consequence, we regain the N/S conductance result when $\chi=0$ instead of $\chi=\pi/2$ as reported in Ref.~\onlinecite{sengupta}. Physically, this seems to be more reasonable since $\chi=0$ corresponds to the case of an absent barrier, a situation where there is no source of normal reflection besides the condition that momentum in the direction parallell to the barrier must be conserved. We have also verified that our $\chi=0$ result coincides with the results obtained using the full expressions (see Appendix) without assuming a thin-barrier limit when we let both $w$ and $V_0$ go to zero. 
We believe that this minor discrepancy between our results and the results of Ref.~\onlinecite{sengupta} stems from a sign error in their Eq. (5) and also in their expression for $k_b(k_b')$ in the text above Eq. (5).
\par
\begin{figure}[h!]
\centering
\resizebox{0.475\textwidth}{!}{
\includegraphics{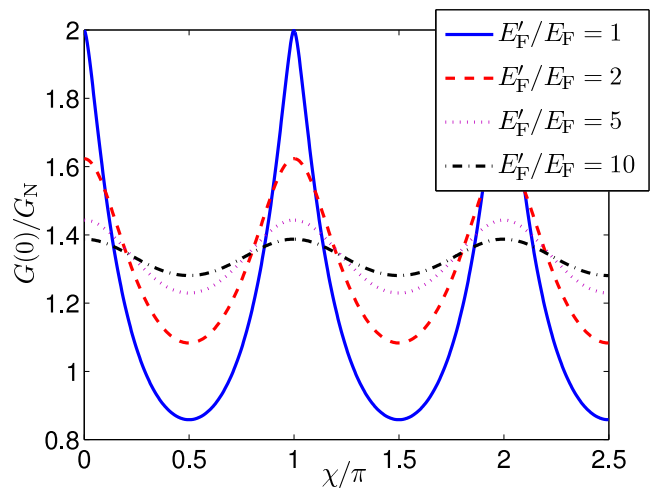}}
\caption{(Color online) Plot of the zero-bias conductance as a function of $\chi$ with $E_\text{F}/\Delta=100$ for a N/I/S graphene junctions with $s$-wave pairing.}
\label{fig:chiswave}
\end{figure}

To unveil the periodicity even more clear, consider Fig. \ref{fig:chiswave} for a plot of the zero-bias conductance as a function $\chi$. The cases $E_\text{F}'=E_\text{F}$ and $E_\text{F}'\neq E_\text{F}$ display a striking difference. The qualitative shape of the curves is equal, but the amplitude is diminished with increasing FVM. This may in similarity to the above discussion be attributed to the increased normal reflection that takes place at zero bias voltage, thus reducing the conductance.
\par

\subsubsection{Anisotropic $d$-wave pairing}
\begin{figure}[h!]
\centering
\resizebox{0.475\textwidth}{!}{
\includegraphics{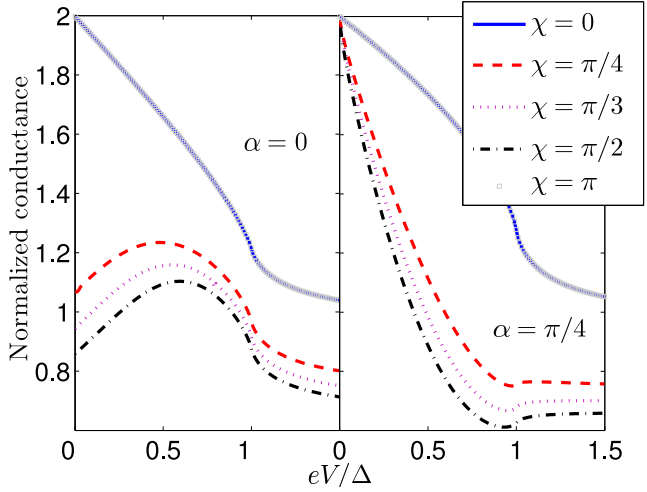}}
\caption{(Color online) Conductance spectra for an N/I/S graphene junction with $E_\text{F}'/\Delta=E_\text{F}/\Delta = 100$, using $d$-wave pairing.}
\label{fig:nisdwave1}
\end{figure}

We now contrast the $s$-wave case with an anisotropic pairing potential to see how the spectra are altered. Consider first Fig. \ref{fig:nisdwave1} for a plot of the tunneling conductance in the undoped case. We consider the two angles $\alpha=0$ and $\alpha=\pi/4$ as representatives for the two types of qualitative behaviour that may be expected in a $d$-wave superconductor/normal graphene junction. The latter corresponds to perfect formation of ZES in the metallic counterpart junction. From the spectra, one infers that for $\alpha=0$, tunneling into the nodes of the superconducting gap destroys the nearly perfect Andreev reflection for subgap energies obtained in Fig. \ref{fig:nisswave}. When $\alpha=\pi/4$, one observes the formation of a ZBCP which peaks at twice the normal state conductance. It is also interesting to note that the zero bias conductance remains unchanged upon increasing $\chi$. Therefore, the equivalent of Fig. \ref{fig:chiswave} in the present $d$-wave case is $G(0)\simeq2G_\text{N}$, regardless of $\chi$.
\par
\begin{figure}[h!]
\centering
\resizebox{0.475\textwidth}{!}{
\includegraphics{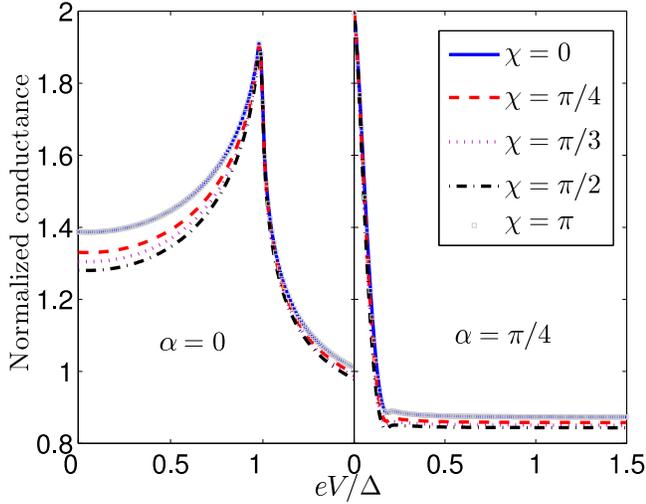}}
\caption{(Color online) Conductance spectra for an N/I/S graphene junction with $E_\text{F}'/E_\text{F} = 10$ and $E_\text{F}/\Delta=100$, using $d$-wave pairing.}
\label{fig:nisdwave2}
\end{figure}

Introducing a FVM between the superconducting and normal parts of the system, the spectra are rendered less sensitive to any increase in $\chi$, as seen in Fig. \ref{fig:nisdwave2}. For $\alpha=0$, the spectra are essentially identical to the doped $s$-wave case. For $\alpha=\pi/4$, it is seen that the formation of a ZBCP becomes even more protruding, and that the zero bias conductance is still insensitive to any increase in $\chi$. Therefore, one is led to conclude that the normalized zero bias conductance $G(0)/G_\text{N}$ in the $d$-wave case is constantly equal to nearly 2, regardless of $\chi$ \textit{and} the magnitude of the FVM.

\section{Conductance-oscillations}
In this section, we investigate a feature of the conductance spectra that is in common for both the $s$-wave and $d$-wave case: an oscillatory behaviour as a function of applied bias voltage. 
Consider first a N/I/S graphene junction. In the 
thin-barrier limit defined as $w\to0$ and $V_0\to\infty$ with $s$-wave pairing, Ref.~\onlinecite{sengupta} 
reported a $\pi$-periodicity of the conductance with respect to the parameter $\chi = V_0w/v_\text{F}$, as discussed in the previous section. We now show that by not restricting ourselves to the thin-barrier limit, 
new physics emerges from the presence of a finite-width barrier. We measure the width $w$
of region I in units of $\lambda_\text{F}'$ and the potential barrier $V_0$ in units of $E_\text{F}'$. 
The linear dispersion approximation is valid \cite{wallace} up to $\simeq 1$ eV, and we will consider 
Fermi energies in graphene \cite{novoselov} ranging from the undoped case $E_\text{F} \approx 0$ meV to $E_\text{F} \approx 100$ meV in the doped case, setting the gap value to
$\Delta = 1$ meV. Owing to the restriction of $E_\text{F}'\gg\Delta$, we fix $E_\text{F}'= 100\Delta$, and also set $V_0 = 500\Delta$ in order to model the effective potential barrier.
\par

\begin{figure}[h!]
\centering
\resizebox{0.5\textwidth}{!}{
\includegraphics{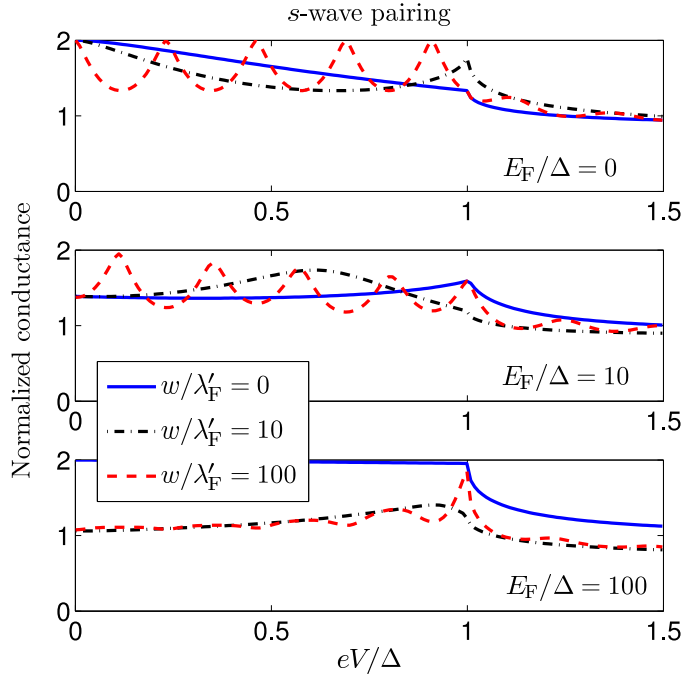}}
\caption{(Color online) Tunneling conductance of N/I/S graphene junction for $s$-wave pairing in the undoped and doped case (see main text for parameter values). We have fixed $V_0/\Delta = 500$ and $E_\text{F}'/\Delta = 100$. It is seen that for increasing $w$, a novel oscillatory behaviour of the conductance as a function of voltage is present in all cases.}
\label{fig:swaveosc}
\end{figure}

\begin{figure}[h!]
\centering
\resizebox{0.5\textwidth}{!}{
\includegraphics{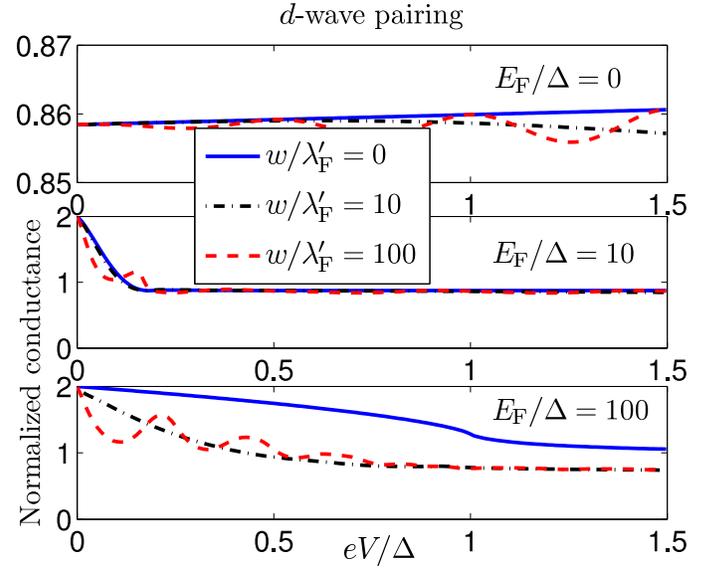}}
\caption{(Color online) Same as Fig. \ref{fig:swaveosc}, but now for $d$-wave pairing. We have fixed $V_0/\Delta = 500$ and $E_\text{F}'/\Delta = 100$.}
\label{fig:dwaveosc}
\end{figure}

Consider Fig. \ref{fig:swaveosc} where we plot the normalized tunneling conductance in case
of $s$-wave pairing, for both a doped and undoped normal part of the system. The most striking new feature 
compared to the thin-barrier limit is the strong oscillations in the conductance as a function of $eV$.
For subgap energies, we regain the N/S conductance for undoped graphene when $\chi=0$, with nearly 
perfect Andreev reflection. The same oscillations are seen in the $d$-wave pairing case, shown in Fig. \ref{fig:dwaveosc}. To model the $d$-wave pairing, we have used the 
$d_{x^2-y^2}$ model $\Delta(\theta) = \Delta\cos(2\theta-2\alpha)$ with $\alpha=\pi/4$. The parameter 
$\alpha$ effectively models different orientations of the gap in $\vk$-space with regard to the 
interface, and $\alpha=\pi/4$ corresponds to perfect formation of ZES in N/S metallic 
junctions. For $\alpha=0$, the $d$-wave spectra are essentially identical to the $s$-wave case, since the condition for formation of ZES is not fulfilled in this case \cite{tanaka}. It is 
seen that in all cases shown in Figs. \ref{fig:swaveosc} and \ref{fig:dwaveosc} the conductance exhibits a novel oscillatory 
behavior as a function of applied bias voltage $eV$ as the width $w$ of the insulating region becomes 
much larger than the Fermi wavelength, \ie $w\gg \lambda_\text{F}'$.
\par 
The oscillatory behavior of the conductance may be understood as follows. Non-relativistic free 
electrons with energy $E$ impinging upon a potential barrier $V_0$ are described by an expontentially 
decreasing non-oscillatory wavefunction $\e{\i kx}$ inside the barrier region if $E<V_0$, since the 
dispersion essentially  is $k\sim\sqrt{E-V_0}$. Relativistic free electrons, on the other hand, have a 
dispersion $k\sim (E-V_0)$, 
such that the corresponding wavefunctions do not decay inside the barrier region. Instead, the transmittance 
of the junction will display an oscillatory behavior as a function of 
the energy of incidence $E$. In general, 
a kinetic energy given by $\sim k^{\alpha}$ will lead to a complex momentum $k \sim (E-V_0)^{1/\alpha}$
inside the tunneling region, and hence damped oscillatory behavior of the wave function. Relativistic 
massless fermions are unique in the sense that only in this case ($\alpha = 1$) is the momentum purely 
real. Hence, the undamped oscillatory behavior at sub-gap energies appears as a direct manifestation
of the relativistic low-energy Dirac fermions in the problem. This observation is also linked to the so-called Klein paradox which occurs for electrons with such a relativistic dispersion relation, which has been theoretically studied in normal graphene \cite{katsnelson}.
\par
We next discuss why the illustrated conductance spectra are different for $s$-wave and $d$-wave symmetry, in 
addition to comparing the doped and undoped case. The difference in doping level between the superconducting and normal part of the system may be considered as an effective FVM, 
acting as a source of normal reflection in the scattering processes. This is why the subgap conductance at 
thin barrier limit is reduced when $E_\text{F}'\neq E_\text{F}$. Moving away from the thin barrier limit, it is seen 
that oscillations emerge in the conductance spectra. For $s$-wave pairing, the amplitude of the oscillations 
is larger for $E_\text{F}'\neq E_\text{F}$ than for the case of no FVM, but the period of oscillations remains the same. This 
period depends on $w$, while the amplitude of the oscillations is governed by the wavevectors in the regions 
I and S. The maximum value of the  oscillations occurs when $2w$ equals an integer number of wavelengths,
corresponding to a constructive interference between the scattered waves. Physically, the amplitude-dependence 
of the oscillations on the doping level originates with the fact that any FVM effectively acts as an 
increase in barrier strength. By making $V_0$ larger, one introduces a stronger source of normal reflection. 
When the resonance condition for the oscillations is not met, the barrier  reflects the incoming particles 
more efficiently. This is also the reason why increasing $V_0$ directly and increasing the FVM has 
the same effect on the spectra.
\par
We now turn to the difference between the $s$-wave and $d$-wave symmetries. It is seen that 
the conductance is reduced in the $d$-wave case compared to the $s$-wave case, and is actually nearly constant for $E_F=0$. One may understand the 
reduction in subgap conductance in the undoped case as a consequence of tunneling into the nodes of the 
gap, which is not present in the $s$-wave case. Hence, Andreev reflection which significantly contributes 
to the conductance, is reduced in the $d$-wave case compared to the $s$-wave case. Moreover, we see that 
a ZBCP is formed when $E_\text{F}'\neq E_\text{F}$, equivalent to a stronger barrier, and this is interpreted as the 
usual formation of ZES leading to a transmission at zero bias with a sharp drop for 
increasing voltage.

\section{Discussion}\label{sec:discussion}
By means of the proximity effect, Heersche \etal \text{ } successfully induced superconductivity in a graphene layer \cite{heersche} (see also Ref.~\onlinecite{du}). This achievement opens up a \textit{vista plethora} of new, exciting physics due to the combination of the peculiar electronic features of graphene and the many interesting properties of superconductivity. For our theory to be properly tested experimentally, it is necessary to create N/S and N/I/S graphene junctions. Junctions involving normal graphene with insulating regions have recently been experimentally realized \cite{novoselov, zhang}. In our work, we have discussed novel conductance-oscillations in a N/I/S graphene junction that arise when moving away from the thin-barrier limit discussed in Ref.~\onlinecite{sengupta}. While reaching the thin-barrier limit might pose some difficulties from an experimental point of view, our predictions are manifested when using wide barriers, which should be technically easier to realize. In order to reach the doped regime, this could be achieved by either chemical doping or using a gate voltage to raise the Fermi level in the superconducting region \cite{katsnelson, milton}. The relevant magnitudes for the various physical quantities present in such an experimental setup has been discussed in the main text of this paper.
\par
It is also worth mentioning that since we have assumed a homogeneous chemical potential in each of the normal, insulating, and superconducting regions, the experimental realization of the predicted effects require charge homogeneity of the graphene samples. This is a challenging criteria, since electron-hole puddles in graphene imaged by scanning single electron transistor \cite{martin} suggest that such charge
inhomogeneities probably play an important role in limiting the transport characteristics of graphene \cite{castro}. In addition, we have neglected the spatial variation of the superconducting gap near the N/S interfaces. The suppression of the order parameter is expected to least pronounced when there is a large FVM between the two regions. However, the qualitative results presented in this work are most likely unaffected by taken into account the reduction of the gap near the interface.

\section{Summary}\label{sec:summary}
In summary, we have studied coherent quantum transport in normal/superconductor (N/S) and normal/insulator/superconductor (N/I/S) 
graphene junctions, investigating also the role of $d$-wave pairing symmetry on the tunneling conductance. We elaborate on the results obtained in Ref.~\onlinecite{linderPRL07}, namely a new oscillatory behaviour of the conductance as 
a function of bias voltage for insulating regions that satisfy $w\gg \lambda_\text{F}'$, which is present both for 
$s$- and $d$-wave pairing. This is a unique manifestion of the Dirac-like fermions in the problem. In the $d$-wave case, we have studied the conductance of an N/S and N/I/S junction in order to make predictions of what could be expected in experiments, providing both analytical and numerical results. We
find very distinct behaviour from metallic N/S junctions in the presence of a FVM: a rotation of $\alpha$ is accompanied by a 
progressive shift of the peak in the conductance, without any formation of a ZBCP except for $\alpha=\pi/4$. All of our predictions should be easily 
experimentally observable, which constitutes a direct way of testing our theory.

\section*{Acknowledgments}
\noindent The authors are indebted to Takehito Yokoyama for very useful comments in addition to critical reading of the manuscript 
and the numerical code, and to Zlatko Tesanovic for helpful comunications. We have also benefited from discussions with 
Annica Black-Schaffer and Carlos Beenakker. This work 
was supported by the Norwegian Research Council Grants No. 158518/431 and No. 158547/431 (NANOMAT), and Grant No. 167498/V30 
(STORFORSK). The authors also acknowledge the Center for Advanced Study at the Norwegian Academy of Science and Letters, for 
hospitality during the academic year 2006/2007. 

\appendix

\section{Normal- and Andreev-reflection coefficients for N/I/S junctions}
Solving the boundary conditions Eq. (\ref{eq:nisboundary}), we obtain the following expressions for the normal reflection coefficient $r$ and the Andreev-reflection coefficient $r_A$:
\begin{align}
r &= t_e(A+C) + t_h(B+D) - 1,\notag\\
r_A &= t_e(A'+C') + t_h(B'+D'),
\end{align}
where the transmission coefficients read
\begin{align}
t_e &= 2\cos\theta[\e{-\i\theta_A}(B'+D') - (B'\e{-\i\tilde{\theta}_A} - D'\e{\i\tilde{\theta}_A})]\rho^{-1},\notag\\
t_h &= t_e[\e{\i\theta_A}(A'\e{-\i\tilde{\theta}_A} - C'\e{\i\tilde{\theta}_A})-A'-C']\notag\\
&\times[B'+D'-\e{\i\theta_A}(B'\e{-\i\tilde{\theta}_A}-D'\e{\i\tilde{\theta}_A})]^{-1},
\end{align}
with the definition
\begin{align}
\rho &= [ \e{-\i\theta_A}(B'+D') - (B'\e{-\i\tilde{\theta}_A} -
D'\e{\i\tilde{\theta}_A})]\notag\\
&\times[ \e{-\i\theta}(A+C) +
(A\e{\i\tilde{\theta}} - C\e{-\i\tilde{\theta}})] \notag\\
&-[(D\e{-\i\tilde{\theta}} - B\e{\i\tilde{\theta}}) -
\e{-\i\theta}(B+D)]\notag\\
&\times[(A'\e{-\i\tilde{\theta}_A}-C'\e{\i\tilde{\theta}_A}) -
\e{-\i\theta_A}(A'+C')]
\end{align}
We have defined the auxiliary quantities
\begin{align}
A &= u_+\e{\i(q^+-p^+)}[1 - (\e{\i\tilde{\theta}} - \e{\i\theta^+})(2\cos\tilde{\theta})^{-1}],\notag\\
B &= v_-\e{\i(q^--p^+)}[1 - (\e{\i\tilde{\theta}} - \e{\i\theta^-})(2\cos\tilde{\theta})^{-1}],\notag\\
C &= u_+\e{\i(p^++q^+)}(\e{\i\tilde{\theta}} - \e{\i\theta^+})(2\cos\tilde{\theta})^{-1},\notag\\
D &= v_-\e{\i(p^++q^-)}(\e{\i\tilde{\theta}} - \e{\i\theta^-})(2\cos\tilde{\theta})^{-1},
\end{align}
and similarly introduced
\begin{align}
A' &= v_+\e{\i(q^++p^- - \phi^+)}[1 + (\e{\i\theta^+} - \e{-\i\tilde{\theta}_A})(2\cos\tilde{\theta}_A)^{-1}],\notag\\
B' &= u_-\e{\i(q^-+p^- - \phi^-)}[1 + (\e{\i\theta^-} - \e{-\i\tilde{\theta}_A})(2\cos\tilde{\theta}_A)^{-1}],\notag\\
C' &= v_+\e{\i(q^+-p^--\phi^+)}( \e{-\i\tilde{\theta}_A}- \e{\i\theta^+} )(2\cos\tilde{\theta}_A)^{-1},\notag\\
D' &= u_-\e{\i(q^--p^--\phi^-)}( \e{-\i\tilde{\theta}_A} -\e{\i\theta^-} )(2\cos\tilde{\theta}_A)^{-1}.
\end{align}
For more compact notation, we have finally defined
\begin{align}
q^+ &= q^\text{e}\cos\theta^+ w,\; q^- = q^\text{h}\cos\theta^- w,\notag\\
p^+ &= \tilde{p}^\text{e}\cos\tilde{\theta} w,\; p^- = \tilde{p}^\text{h}\cos\tilde{\theta}_A w.
\end{align}
In the thin-barrier limit defined as $w\to 0$ and $V_0\to\infty$, one may set
\begin{align}
\tilde{\theta} \to 0,\; \tilde{\theta}_A \to 0,\; q_\pm\to 0,\; p_\pm \to \mp\chi,
\end{align}
where $\chi = V_0w/v_\text{F}$.


\begin{thebibliography}{99}

\bibitem{deutscher} G. Deutscher, Rev. Mod. Phys. \textbf{77}, 109 (2005).

\bibitem{beenakker} C. W. J. Beenakker, Phys. Rev. Lett. \textbf{97}, 067007 (2006).

\bibitem{beenakkerreview} C. Beenakker, arXiv:0710.3848v1 (2007).

\bibitem{novoselov}  K. S. Novoselov, A. K. Geim, S. V. Morozov, D. Jiang, Y. Zhang, S. V. Dubonos, I. V. Grigorieva, A.A. Firsov, Science \textbf{306}, 666 (2004).

\bibitem{zhang} Y. Zhang, Y.-W. Tan, H. L. Stormer, P. Kim , Nature \textbf{438}, 201 (2005).

\bibitem{QED3} 
In the quantum domain at low temperatures, and in the vicinity of the edge of the superconducting 
dome, the coupling between the nodal fermions in high-$T_c$ superconductors  and quantum critical 
phase-fluctuations of the superconducting order parameter (i.e. vortices) will lead to unusual 
behavior. This has important ramifications for constructing a viable theory of the so-called
pseudo-gap phase of these systems. See
M. Franz and Z. Tesanovic, Phys. Rev. Lett. {\bf 87} 257003 (2001);
Z. Tesanovic, O. Vafek, and M. Franz, Phys. Rev B {\bf 65}, 180511 (2002);
O. Vafek and Z. Tesanovic, Phys. Rev. Lett. {\bf 91}, 237001 (2003).
 

\bibitem{heersche}  H. B. Heersche, P. Jarillo-Herrero, J. B. Oostinga, L. M. K. Vandersypen, A. F. Morpurgo, Nature \textbf{446}, 56 (2007)

\bibitem{du} X. Du, I. Skachko, E. Y. Andrei, arXiv:0710.4984.

\bibitem{kasumov} A. Yu. Kasumov, R. Deblock, M. Kociak, B. Reulet, H. Bouchiat, I. I. Khodos, Yu. B. Gorbatov, V. T. Volkov, C. Journet, M. Burghard, Science \textbf{284}, 1508 (1999).

\bibitem{morpurgo} A. F. Morpurgo, J. Kong, C. M. Marcus, H. Dai, Science \textbf{286}, 263 (1999).
\bibitem{buitelaar}  M. R. Buitelaar, W. Belzig, T. Nussbaumer, B. Babic, C. Bruder, C. Schoenenberger, Phys. Rev. Lett. \textbf{91}, 057005 (2003)

\bibitem{jariloo} P. Jarillo-Herrero, J. A. van Dam, and L. P. Kouwenhoven, Nature (London) \textbf{439}, 953 (2006).

\bibitem{sengupta} S. Bhattacharjee and K. Sengupta, Phys. Rev. Lett. \textbf{97}.


\bibitem{mazinjohannes} I. I. Mazin and M. D. Johannes, Nat. Phys. \textbf{1}, 91 (2005).

\bibitem{gonzales1999} J. Gonzàlez, F. Guinea, M. A. H. Vozmediano, Phys. Rev. B \textbf{59},
R2474 (1999).

\bibitem{kane2004} C. L. Kane and E. J. Mele, Phys. Rev. Lett. \textbf{93}, 197402 (2004).

\bibitem{novoselov_nature} K. S. Novoselov, A. K. Geim, S. V. Morozov, D. Jiang, M. I. Katsnelson,
I. V. Grigorieva, S. V. Dubonos, and A. A. Firsov, Nature \textbf{438}, 197 (2005).

\bibitem{linderPRL07} J. Linder and A. Sudb{\o}, Phys. Rev. Lett. \textbf{99}, 147001 (2007).

\bibitem{hu} C.-R. Hu, Phys. Rev. Lett. \textbf{72}, 1526 (1994).

\bibitem{tanaka} Y. Tanaka and S. Kashiwaya, Phys. Rev. Lett. \textbf{74}, 3451 (1995).

\bibitem{wallace} P. R. Wallace, Phys. Rev. \textbf{71}, 622 (1947).

\bibitem{bt} G. E. Blonder and M. Tinkham, Phys. Rev. B \textbf{27}, 112 (1983).

\bibitem{zutic} I. Zutic and O. T. Valls, Phys. Rev. B \textbf{61}, 1555 (2000). 

\bibitem{kashiwaya96} S. Kashiwaya, Y. Tanaka, M. Koyanagi, K. Kajimura, Phys. Rev. B \textbf{53}, 2667 (1996).

\bibitem{bruder} C. Bruder, Phys. Rev. B \textbf{41}, 4017 (1990).

\bibitem{blackschaffer} A. Black-Schaffer and S. Doniach, Phys. Rev. B \textbf{75}, 134512 (2007).

\bibitem{sudbo} K. Fossheim, A. Sudb{\o}, {\it Superconductivity: Physics and applications}, John Wiley \& Sons Ltd., Ch. 5 (2004).


\bibitem{kleinert} H. Kleinert, Phys. Rev. Lett. \textbf{84}, 286 (2000).

\bibitem{phase_fluctuations} Z. Tesanovic, Phys. Rev. B {\bf 51}, 16204 (1995); {\it ibid}, B {\bf 59}, 6449 (1999);
A. K. Nguyen and A. Sudb{\o}, Phys. Rev. B {\bf 57}, 3123 (1998); {\it ibid}, B {\bf 58}, 2802 (1998);
{\it ibid}, B {\bf 60}, 15307 (1999); Europhys. Lett. {\bf 46}, 780 (1999). 

\bibitem{RevModPhys_2005} F. S. Bergeret, A. F. Volkov, and K. B. Efetov, Rev. Mod. Phys.
\textbf{77}, 1321 (2005); A. I. Buzdin, Rev. Mod. Phys. \textbf{77}, 935 (2005).


\bibitem{2gas} A. F. Volkov, P. H. C. Magnee, B. J. van Wees, and T. M. Klapwijk, Physica C \textbf{242}, 261 (1995).

\bibitem{wire} G. Fagas, G. Tkachov, A. Pfund, and K. Richter, Phys. Rev. B \textbf{71}, 224510 (2005).

\bibitem{btk} G. E. Blonder, M. Tinkham, and T. M. Klapwijk, 
Phys. Rev. B {\bf 25}, 4515 (1982).

\bibitem{sengupta2} S. Bhattacharjee, M. Maiti, K. Sengupta, arXiv:0704.2760.

\bibitem{katsnelson} M. I. Katsnelson, K. S. Novoselov, A. K. Geim, Nature Phys. \textbf{2}, 620 (2006)

\bibitem{milton} J. Milton Pereira Jr., P. Vasilopoulos, and F. M. Peeters, cond-mat/0702596.

\bibitem{martin} J. Martin, N. Akerman, G. Ulbricht, T. Lohmann, J. H. Smet, K. von Klitzing, and A. Yacoby, cond-mat/0705.2180 (2007).

\bibitem{castro} E.-A. Kim and A. H. Castro Neto, arXiv:cond-mat/0702562 (2007).


\end{thebibliography}
\end{document}